# Laser based optical interferometer manometer design for primary pressure standard in India

Manoj Das[1,2,*], Sandip Kumar Ghosh [1,2], Kuldeep Kumar[1,2], Elizabeth Jeessa James[2,3], Megha Singh[3] and Ashok Kumar[2,3,*]

(1) Time & frequency Metrology, Indian Standard Time Division, CSIR-National Physical Laboratory, New Delhi
(2) Academy of Scientific and Innovative Research (AcSIR), Ghaziabad 201002
(3) Pressure, Vacuum & Ultrasonic Metrology, Physico-mechanical Division, CSIR-National Physical Laboratory, New Delhi

*e-mail: manoj.das@nplindia.org & ashok553@nplindia.org

**Abstract**

The SI unit of pressure, i.e. Pascal is realized with mechanical devices such as ultrasonic interferometer manometers and pressure balances for the pressure range 1 Pa to 100 kPa. Recently, optical interferometer manometers are being used to realize Pascal. Such a realization is mercury-free and environmentally friendly and does not depend on any mechanical motions as in piston gauges and is based on physical constants. It consists of an optical Fabry-Perot cavity as a refractometer and is based on the measurement of the resonant frequency of the cavity using lasers. In this paper, we report the theoretical calculations for such an optical cavity and discuss various parameters for the laser required for developing such a cavity-based refractometer system. We present the mechanical design for the dual cavity and an all-fiber optical set-up to be used for the next generation of primary pressure standards in the barometric region of pressure at CSIR-National physical laboratory, India.

**1. Introduction**

CSIR-National physical laboratory India (NPLI) is the designated national metrology institute (NMI) of India having the mandate of generation and dissemination of pressure SI unit Pascal and maintaining traceability with the international standards body, BIPM [1]. NPLI has been maintaining primary pressure standards based on ultrasonic interferometer manometers (UIM) [2] and piston gauges [3], for the pressure range 1 Pa to 350 kPa. In 2019, the SI units convention agreed that all SI units are to be defined in terms of fundamental constants, and the value of the Boltzmann constant was fixed [4]. This not only provides long-term stability to the SI units but also opens up the opportunity for the use of quantum technologies to implement the new definition. The present method of UIM and gauges being dependent on toxic mercury and mechanical action respectively, alternative routes to realize the Pascal is being vigorously pursued around the world [5]. A cell-based refractometer has been used to determine the density of helium by measuring the optical refractivity and hence pressure using the equation of state [6]. Another widely researched method is based on optically measuring the gas number density using an optical Interferometer manometer and is in various stages of implementation in NMI's like NIST (USA), NIM (China), PTB (Germany), NMIJ (Japan), KRISS (South Korea) [7–10]. It is based on optical Fabry-Perot (FP) cavities, which effectively measure the refractivity of the gas particles based on the measurement of resonant frequencies. The gas density is calculated from refractivity [7]. The final pressure can be derived from gas density and measurement of the temperature of the system. The dependence of the method on the measurement of frequencies is important, as frequency is the most accurately measurable physical quantity [11].

At NPLI, we are developing a pressure standard using an optical Interferometer manometer, based on fixed length optical cavity. The system consists of a dual cavity [12] where one will be at vacuum (the reference) and the other will be at test gas pressure. The resonant frequency difference between the two cavities will be a measure of the pressure of the gas. In this paper, we report on the optical physics package for the dual FP cavity system. It involves theoretical estimations of the cavity parameters and the ready-to-be-fabricated mechanical layout of the dual cavity. We also discuss the various laser-related requirements and present an all-fiber optical setup for the measurement of the resonant frequencies for minimum day-to-day alignment requirements. This should result in long-term reliable operation, a pre-requisite for any primary standard.

In section 2, we discuss the basic theoretical and experimental methodology. In section 3, we present the results of the theoretical calculations and the mechanical design for the NPL dual cavity. We discuss the requirements regarding the laser, the locking technique, and an all-fiber optical setup.

## 2. Method

Mercury-based liquid column manometers, which are widely being used in various NMI's as a primary pressure standard (1 Pa to 100 kPa) depend on the measurement of the column height $h$, local gravity $g$, and the fluid density $\rho$, according to the equation of pressure, $p = \rho g h$. The most accurate determination of column height $h$ is carried out using ultrasonic interferometric manometers (UIM) [2, 13]. The UIM standard uses mercury, which is environmentally toxic and has been banned or has been on the list of restricted use by various countries around the world [14]. The measurement with UIM takes time for the mercury to settle down, is vibration sensitive and the device itself has a large physical footprint.

Recently, various NMIs around the world have started working on developing a pressure standard in the range of <100 kPa regime utilizing a laser refractometer for pressure measurement of the test gas. It uses the ideal gas equation, $p = \rho_N K_B T$, where $\rho_N$ is the gas density, $K_B$ is the Boltzmann constant and $T$ is the temperature. In the SI standard brochure in 2019 [4], it was decided that $K_B$ will be fixed to a certain value and Kelvin can be measured with high accuracy. The gas density parameter $\rho_N$ can be measured with high accuracy using laser-based optical interferometry technique, where measurement of change in optical length of a Fabry-Perot (FP) cavity in the test gas pressure gives an estimate of the refractive index and hence the gas number density [12].

Optical FP cavities consist of usually a spacer and two mirrors attached to it. The frequencies of the longitudinal cavity modes depend on the cavity optical length ($n \cdot L_c$), where $n$ is the refractive index of the test gas medium and $L_c$ is the physical length. When a laser is locked to a specific resonance of the FP cavity, the frequency $f_l$ of the laser is given by, $f_l \sim \frac{mc}{2nL_c}$, where $m$ is the cavity mode number, and c is the speed light speed. For a change in refractive index $n$ due to applied pressure, the optical length changes and so does the cavity mode positioning. Hence, keeping cavity length changes to a minimum, changes in the refractive index are derived from frequency change $\Delta f$ measurements in $f_l$.

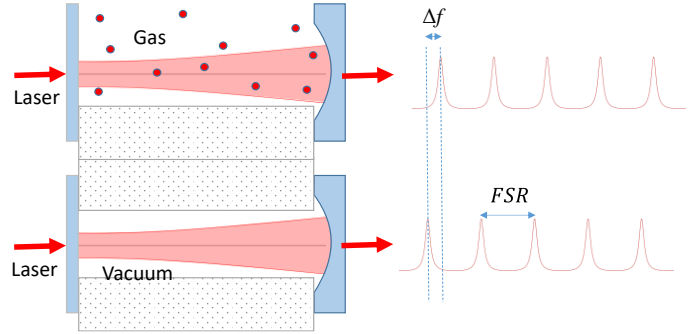

Fig. 1: Schematic of the basic principle of operation for the dual cavity. The transmission longitudinal cavity modes are shown. The shift ($\Delta f$) in longitudinal mode frequency is due to the test gas pressure of the measurement cavity. FSR is the free spectral range of the cavity.

Figure 1 shows the schematic for the dual cavity-based measurement. One of the cavities is at vacuum and serves as the reference cavity and the other cavity (measurement cavity) is filled up with the gas whose pressure is to be investigated. Two separate lasers are locked to the resonant frequency of the cavities and their absolute frequencies and relative beat note frequencies are observed. First, both the cavities, which share the same spacer and the mirrors, are maintained at vacuum and the rf beat frequency $f_{B1}$ between the resonances of the two cavities is measured. Then the measurement cavity is exposed to the test gas pressure and a final RF beat frequency, $f_{B2}$ is measured. $f_R$ is the resonant frequency of the cavity at the filled-up pressure, $FSR$ is the free spectral range of the measurement cavity and $\varepsilon$ is the cavity mirror dispersion. The frequency difference between the lasers locked to adjacent cavity modes gives the FSR of the cavity. The mirror dispersion information is obtained from the mirror coating manufacturer. The fractional change in resonance frequency is expressed as

$$\frac{\Delta f}{f} = FSR(1+\varepsilon)\frac{\frac{(f_{B1}-f_{B2})}{FSR}+\Delta m}{f_{B2}+f_R} \qquad (1)$$

when pressure is applied in the measurement cavity, the beat frequency increases, and when it is more than the FSR of the cavity, the change in cavity mode number, Δm has to be taken into account. The gas density, $\rho_N$ is derived from refractivity by using the equation of state (ideal gas law) and the Lorentz-Lorenz equation [7]. The final pressure is then calculated from the expression [7, 13]

$$P = A\left(\frac{\Delta f}{f}\right) - B\left(\frac{\Delta f}{f}\right)^2 + C\left(\frac{\Delta f}{f}\right)^3 \quad (2)$$

where A, B, and C are coefficients that depend on distortion due to the pressure for the measurement and reference cavity, as well as on refractivity virial coefficient, density virial co-efficient, Boltzmann constant and thermodynamic temperature. The coefficients have been estimated in Ref. [13].

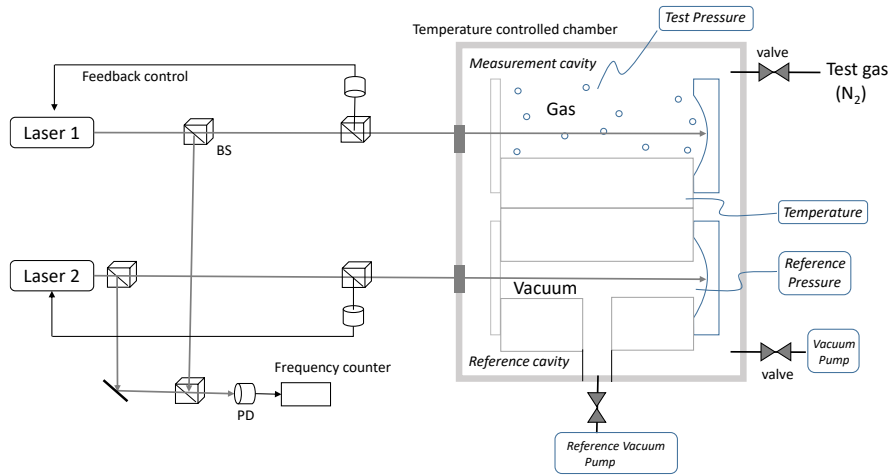

Fig. 2: Schematic of the basic methodology of the experiment. BS: Beam splitter; PD: Photodetector. The locking scheme is not shown.

Figure 2 shows the schematic of the experimental setup showing the basic methodology. Two single-frequency tunable lasers are locked to the cavities independently. The reference cavity is in vacuum and the measurement cavity contains the test gas. The difference in frequency between the two locked lasers, which is the indicator of the change in the pressure due to the test gas, is measured using a frequency counter. The optical length of the cavity changes not only due to the pressure but also from the changes in the ambient temperature. So, the dual cavity should be kept in a temperature-controlled environment. Arrangements for absolute laser frequency measurements are needed for the estimation of the test pressure (Eq. 1). This is usually carried out with a wavemeter referenced to a standard traceable wavelength.

The lasers are locked to the cavity resonance using Pound-Drever-Hall (PDH) technique [15]. The details are in section 3.

## 3. Result: Design and Construction

In this section, we present the cavity and its transverse and longitudinal mode characteristics. We discuss the material property of the mirrors and the spacer required to minimize cavity-induced noises in precise measurement of the resonant frequencies. We present a detailed requirement regarding the lasers and associated components and an all-fiber optical experimental layout.

The FP cavity design wavelength is around 1542 nm, at the center of the telecom wavelength range, owing to the availability of a wide range of optical components and lasers. As the manometer will be used for metrology purposes, comparison of test measurements with peer NMIs is an essential part. So, a similar design approach as that adopted by other leading NMIs has been taken for the cavity as well as the lasers.

### A. Dual optical cavity

Theoretical calculations

An identical two-mirror FP cavity is used. A finesse of $\mathcal{F}_{Cavity} \sim 10,000$, for the cavity, has been considered. A moderate finesse has been chosen so that the mirror surface or coating will not deteriorate significantly when exposed to the test pressure environment [12]. The finesse is $\mathcal{F}_{Cavity} \approx \pi\sqrt{R}/(1-R)$ for mirrors having the same reflectivity, $R$ (mirror surface inside the cavity) [16]. Figure 3(a) shows the plot of finesse with respect to reflectivity. The calculated mirror reflectivity for realizing the above finesse is $> 99.97\%$. Mirrors with these parameters are readily available without the need for costly custom coating runs. The relative cavity length fluctuations determine the fluctuations in frequency ($\Delta f$) of the locked laser, $\Delta f/f \sim \Delta L/L_c$ [17, 18], where $\Delta L$ is the length fluctuations due to temperature & pressure and $L_c$ is the cavity length. A cavity length of $L_c$=150 mm is chosen for our design, for ease of measurement comparisons with other NMIs [19]. The free spectral range is, $FSR = c/2nL_c = 1\ GHz$ [Fig. 3(b)]. The linewidth of the cavity as given by $\Delta\nu_{Cavity} = FSR/\mathcal{F}$ is 100 kHz.

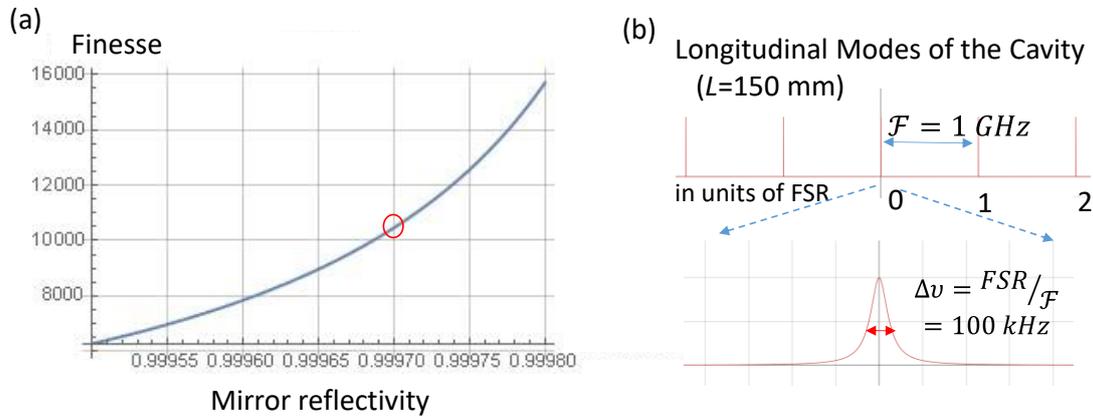

Fig. 3: (a) The graph showing finesse vs mirror reflectivity, The red circle represents the finesse considered; (b) Plot of the longitudinal modes of the cavity in units of FSR (upper); Expanded view of each longitudinal mode

The two mirrors of the FP cavity are plano/plano ($M_1$) and plano/concave ($M_2$) mirrors (Fig. 4(a)). A mirror has two surfaces. Mirror $M_1$ is flat on both sides and mirror $M_2$ is flat on the outside and concave on the inside of the cavity. Such an $M_1$ - $M_2$ combination allows stable cavity operation. Both cavities have identical 2-mirror properties like reflection and transmission. The cavity should support only one stable transverse $TEM_{00}$ mode. The concave mirror RoC of 500 mm has been chosen to support a Gaussian mode which has its waist at the plane mirror position and its wavefront at 150 mm matches with the RoC of $M_2$ [Fig. 4(a)].

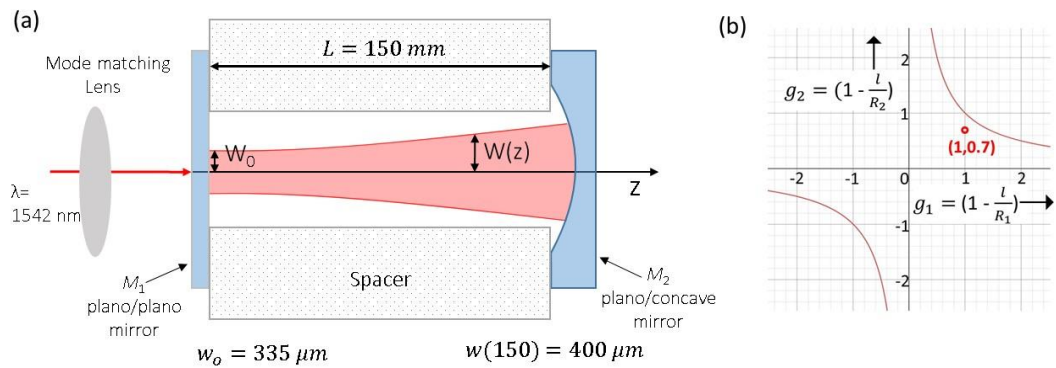

Fig. 4: (a) Schematic of the $TEM_{00}$ mode propagation in the $z$-direction. Mode size parameters of the cavity are shown at the $M_1$ and $M_2$ mirror position; (b) Stability plot of the cavity. The red circle shows the designed stability point. The region bounded by the solid curve shows the stable region

To check whether the cavity modes are oscillating in stable conditions, the stability equation $0 \leq g_1 g_2 \leq 1$ is shown in Fig. 4(b). The parameter, $g_1 = 1 - L_c/R_{M1}$, and $g_2 = 1 - L_c/R_{M2}$, where $R_{M1} = \infty$ is the radius of curvature (RoC) for plane mirror $M_1$, and RoC for $M_2$, $R_{M2} = 500$ mm. The stability factor is $(g_1, g_2) \equiv (1, 0.7)$, as shown by the solid red circle in Fig. 4(b), and lies in the stable region bounded by the solid curves [15]. A suitable mode-matching optics can be used to excite only the stable single $TEM_{00}$ cavity mode. The Gaussian beam size of the TEM$_{00}$ mode is given by, $\omega(z) = \omega_0\sqrt{1 + (z/z_R)^2}$ [16], where Rayleigh range, $z_R = \pi{\omega_0}^2/\lambda$, $\omega_0$ = beam Waist, $\lambda = 1542\ nm$, and $z$ is the propagation direction. The beam size is, $\omega_0 \sim 335\ \mu m$ and $\omega(150\ mm) \sim 400\ \mu m$ [Fig. 4(a)]. Depending on the input beam size, a proper focal length lens (mode matching lens) can generate the required waist size at the $M_1$ mirror position, to trigger the designed $TEM_{00}$ mode inside the cavity.

Mechanical Design and Material Considerations

The applied pressure causes a change in the optical length of the cavity through the refractive index, $n$, assuming that the physical length $L$ of the cavity is constant. The length of the spacer determines the cavity length, $L$. Any fluctuations in the spacer length will have a direct impact on the pressure measurement. Low-expansion glasses are used for realizing such spacers. A typical order of magnitude coefficient of thermal expansion (CTE) of better than $\pm 10$ ppb/$^0$C in the temperature range 290 K to 305 K is preferred for the spacer with zero crossing temperature around room temperature. Figure 5(a) shows the design for the dual cavity at CSIR-NPL.

The dual FP cavity is designed to be made from a single block of spacer material with two channels. A single block as a spacer is used to cancel out common frequency shifts due to the changes in the length of the spacer. Both the channels are cylindrical in shape with a vacuum hole in one and a cutout slot in the other [Fig. 5(a)]. The two pairs of mirrors are optically bonded onto the spacer end faces across the channels creating a dual cavity system. The cavity with a hole is connected to a vertical tube and then to a metal flange (not shown in the figure) for vacuum piping and acts as the reference cavity. The other cavity with the slot has one side exposed to the environment enclosing the cavity, to allow the flow of the test gas into the channel, and is the measurement cavity. Figure 5(b) shows the schematic and the mechanical dimensions of the dual cavity.

Both the spacer and the cavity have been chosen to be constructed from ULE® glass [20]. The selection of material for the spacer and the mirrors decides the performance of the manometer. All the bonding between the mirror and the spacer and between the spacer and the vent are optical contacts to avoid the use of adhesives, which will have different thermal expansion properties. The curved mirrors have an annulus (an outer ring of flat surface for contact). The adhesives will also have some outgassing which will degrade the environment and the overall pressure measurements.

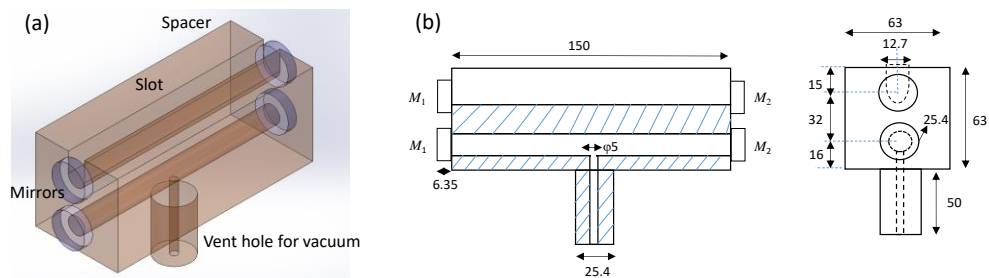

Fig. 5: (a) 3D CAD diagram of the proposed NPLI dual cavity OIM (material: ULE); (b) Mechanical drawings. Units are in mm.

The loss (absorption, scattering) of the mirror materials should be less than the transmission through the mirrors. The details are explained in the lasers section. The designed transmission for the mirrors is 200 ppm (0.02%) and the loss is less than 100 ppm. The outer surface of the mirrors is coated for anti-reflection (<0.2%) and the inner surface for high reflectivity (>99.97%).

**B. Lasers and all-fiber optical set-up**

The lasers play a crucial part in the OIM. Two separate lasers are used to probe the cavities. An external cavity diode laser (ECDL) has been chosen with its fast and slow frequency tuning properties and relatively large mode hop-free range (> 20 GHz)(see Fig. 6) [9, 21]. Such lasers consist of usually a grating, which acts as a frequency filter as well as a mirror for the external cavity. They provide a means to tune the frequency to bandwidths of 10s of kHz. The current supplied to the laser diode itself can also be modulated to achieve faster feedback of the order of 1 MHz. After an optical isolator, the laser output is coupled to a polarization-maintaining single-mode fiber. The optical arrangement for both lasers is the same. There are three functions to be performed. One is the wavelength measurement for each laser. The second is locking to the cavity mode. And third, is the difference in frequency measurement between the two lasers locked independently to the cavities. Mostly, such optical arrangement is realized using free-space optics. But, such an arrangement is prone to day-to-day alignment issues and hampers long time reliable operation of the OIM. We have designed the optics with an all-fiber optical setup. The wavelength being 1542 nm, all the fiber-based optics at telecom wavelength can be readily used.

The light from the ECDL output is first split with a 99:1 fiber splitter. These are commercial single-mode fused fiber optic couplers with standard readily available coupling ratios. The smaller part is used for the wavemeter and the rest is then again split in a 90:10 ratio. After the second splitter, the maximum power is sent to the laser locking unit, and the remaining power is used to provide a signal for the beat signal detection unit. The beat signal detection unit is an all-fiber optical unit with a combiner for mixing the samples from the two lasers and then a polarization controller to sample that part of the light from the two lasers which are lasing in the same polarization. A fast photodiode (~9 GHz) is used to detect the signal and the output is fed to a frequency counter. The fiber-based optical arrangement provides robustness and brings flexibility in using the OIM outside the laboratory environment. The part which is earmarked for locking to the cavity is brought to the dual cavity location and collimated into free space using a fiber collimator.

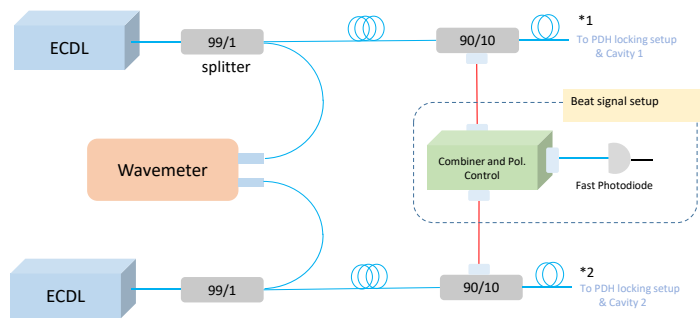

Fig. 6: Experimental realization of the all-fiber optical layout for the laser. Splitters 99:1 and 90:10 can be seen to split the power. The notation (*1): fiber to the cavity location as shown in Fig. 7. (*2) is a similar arrangement as (*1).

Figure 7 shows the locking arrangement of the cavity using the PDH technique [15]. The fiber-collimated output light is sent to an electro-optic modulator (EOM). The EOM modulates the carrier input frequency by 25 MHz to create side-bands, required to generate the error signal. The EOM light output is mode matched to a $TEM_{00}$ mode of the cavity using a lens. The back reflected light from the cavity mirror $M_1$ is picked up by a polarizing beam splitter (PBS) with a help of a quarter wave plate ($\lambda/4$) and sent to the fast photodetector (PD, reflection detector). The light transmitted through the cavity is monitored by a large-area photodetector. The output of the reflection detector is sent to a mixer, where it is demodulated with the EOM 25 MHz (phase shifted) signal and the output is used as an error signal. A suitable feedback signal through a slow and fast channel is directed to the ECDL grating-PZT feedback and the current feedback respectively.

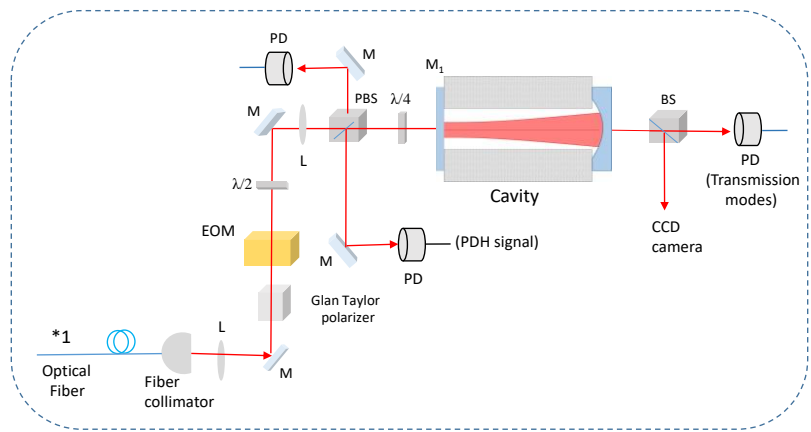

Fig. 7: The PDH technique details for locking the ECDL to the longitudinal mode of one of the cavities. (*1) from Fig. 6. EOM: Electro-optic modulator; PD: photodiode; M: mirror; PBS: polarizing beam splitter; $\lambda/4$: Quarter wave plate; CCD: Charge-coupled device

Two PDH setup is used to lock the lasers to the dual cavity independently. Once the lasers are locked to the longitudinal mode of the cavity, their beat frequency measurement will be then carried out using a frequency counter. Typically, for nitrogen a pressure change of 1 Pa corresponds to a frequency change of ~1.3 MHz [9].

The transmission of the mirrors plays an important role in determining the PDH error signal response. The reflectivity (R) of the mirrors should not be made very high (to increase the $\mathcal{F}_{Cavity}$ and decrease the cavity linewidth), as this will decrease the transmission (T) ($R + T + Loss = 1$, as some amount of loss will be always present due to scattering and absorption). The PDH error signal response depends on $T$ of the incident mirror $M_1$ of the cavity, for a given finesse [22].

## 4. Conclusion

We report on the design for the next-generation primary pressure standard based on the dual optical cavity, for use in the 1Pa to 100 kPa range. We presented the theoretical calculations for the optical cavity and discuss various parameters for the lasers required for developing such a cavity-based pressure measurement system. We present the mechanical design for a dual cavity and an all-fiber optical set-up, for low maintenance long-term reliable operation. This work paves the way for the construction of a primary pressure standard in India. This is in line with the recommendations of the BIPM to develop standards based on physical constants and should complement the pressure measurements of other metrology institutes using such a dual cavity system.

## 5. Acknowledgement

Sandip Kumar Ghosh and Kuldeep Kumar want to acknowledge the University Grants Commission (UGC) and Council of Scientific and Industrial Research (CSIR) respectively for fellowship support to carry out this work. Manoj Das wants to thank the Director (NPL) and Head (Indian standard time division) for their constant support and encouragement. Ashok Kumar wants to acknowledge the Council of Scientific and Industrial Research (CSIR) for funding this project (MLP210532).